\documentclass[sigconf]{acmart}

\AtBeginDocument{%
  }

\setcopyright{acmcopyright}
\copyrightyear{2018}
\acmYear{2018}
\acmDOI{XXXXXXX.XXXXXXX}

\usepackage{siunitx}
\usepackage{xcolor,colortbl}
\usepackage{amsmath}

\begin{document}

\title[picoRing]{picoRing: battery-free rings for subtle thumb-to-index input}


\author{Ryo Takahashi}
\affiliation{%
  \institution{Meta Inc., Reality Labs Research}
  \city{Redmond, Washington}
  \country{USA}
  }
\email{ryotakahashi@meta.com}
\email{takahashi@akg.t.u-tokyo.ac.jp}

\author{Eric Whitmire}
\affiliation{%
  \institution{Meta Inc., Reality Labs Research}
  \city{Redmond, Washington}
  \country{USA}
  }
\email{ewhitmire@meta.com}

\author{Roger Boldu}
\affiliation{%
  \institution{Meta Inc., Reality Labs Research}
  \city{Redmond, Washington}
  \country{USA}
  }
\email{rboldu@meta.com}

\author{Shiu Ng}
\affiliation{%
  \institution{Meta Inc., Reality Labs Research}
  \city{Redmond, Washington}
  \country{USA}
  }
\email{shiung@meta.com}

\author{Wolf Kienzle}
\affiliation{%
  \institution{Meta Inc., Reality Labs Research}
  \city{Redmond, Washington}
  \country{USA}
  }
\email{wkienzle@meta.com}

\author{Hrvoje Benko}
\affiliation{%
  \institution{Meta Inc., Reality Labs Research}
  \city{Redmond, Washington}
  \country{USA}
  }
\email{benko@meta.com}

\renewcommand{\shortauthors}{Ryo Takahashi et al.}

\begin{abstract}
Smart rings for subtle, reliable finger input offer an attractive path for ubiquitous interaction with wearable computing platforms. 
However, compared to ordinary rings worn for cultural or fashion reasons, smart rings are much bulkier and less comfortable, largely due to the space required for a battery, which also limits the space available for sensors.
This paper presents picoRing, a flexible sensing architecture that enables a variety of \textit{battery-free} smart rings paired with a wristband. 
By inductively connecting a wristband-based sensitive reader coil with a ring-based fully-passive sensor coil, picoRing enables the wristband to stably detect the passive response from the ring via a weak inductive coupling. 
We demonstrate four different rings that support thumb-to-finger interactions like pressing, sliding, or scrolling.
When users perform these interactions, the corresponding ring converts each input into a unique passive response through a network of passive switches.
Combining the coil-based sensitive readout with the fully-passive ring design enables a tiny ring that weighs as little as \SI{1.5}{\g} and achieves a \SI{13}{\cm} stable readout despite finger bending, and proximity to metal.
\end{abstract}



\begin{teaserfigure}
  \includegraphics[width=\textwidth]{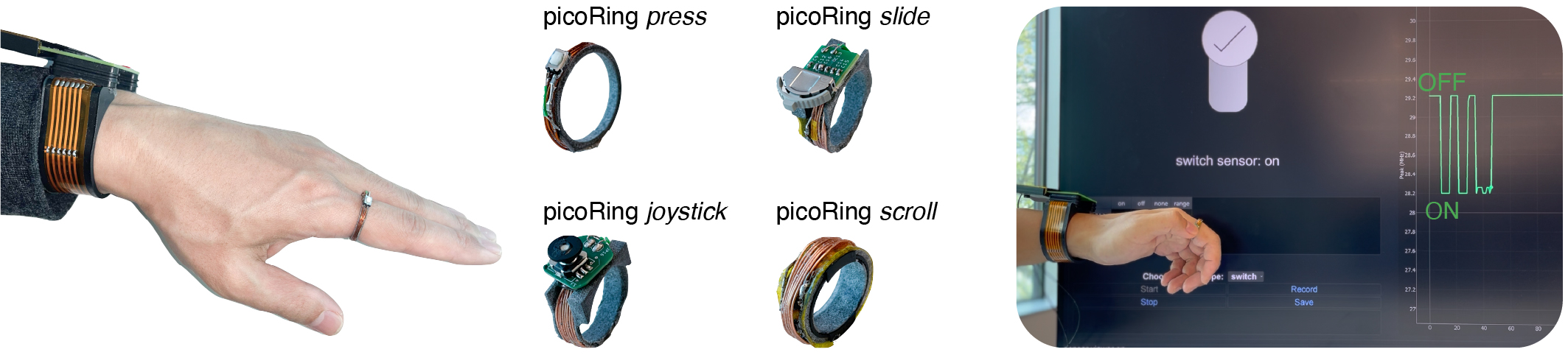}
  \caption{Overview of picoRing. picoRing is a flexible sensing architecture enabling a variety of \textit{battery-free} smart rings paired with a wristband. This paper shows four types of picoRing that support thumb-to-index finger pressing, sliding, or scrolling.}
  \Description{The figure presents a sequence of images displaying a wearable smart device, labeled as 'picoRing', being used for different interactive functions. On the far left, the device is shown worn on a person's wrist, indicating its wearable nature. Moving right, the device is depicted in four variations with labels: 'picoRing press', 'picoRing slide', 'picoRing joystick', and 'picoRing scroll', highlighting different models or interaction modes. Each model is visually distinct and suggests various input methods such as pressing, sliding, or manipulating a joystick. On the far right, the device is shown being used to interact with a digital interface, with a graphic representation of the command output shown as a signal graph that turns a system from "OFF" to "ON".}
  \label{fig:overview}
\end{teaserfigure}

\received{20 February 2007}
\received[revised]{12 March 2009}
\received[accepted]{5 June 2009}

\maketitle

\section{Introduction}

Hand and finger gestures are the primary ways to intuitively and seamlessly interact with wearable computing platforms, such as those provided by smart watches or smart glasses~\cite{Rekimoto2001GestureWrist, Vardy1999TheDevice, Zhai1996The}.
In particular, thumb-to-index finger microgestures support efficient and privacy-preserving inputs for wearable interaction~\cite{Balakrishnan1997Performance,Chan2016User,Vatavu2023IFAD}.
Despite technical advances in wristband-based hand sensing~\cite{Artem2014WristFlex, Chenhan2023aofinger} and glasses-based computer vision~\cite{Grossman2015Typing,Gugenheimer2016FaceTouch}, high-fidelity detection and classification of the subtle and occluded thumb-to-index gestures including tap, pinch, and swipe remains challenging~\cite{Jiang2022Survey, Kienzle2021ElectroRing, Chenhan2023aofinger}.
Some work has explored thumb-to-index finger inputs based on wristband sensing~\cite{Chenhan2023aofinger, Xu2022enabling}, but they require relatively dynamic gestures for reliable classification or limit the input interface to either within-user, short-term usage for sufficient accuracy.
By contrast, a ring-based device with physical inputs, like a button or slider, placed on the index finger offers a simple alternative with unparalleled reliability for subtle thumb-to-index gestures~\cite{Ashbrook2011Nenya, Aakar2019RotoSwype}. 
Such rings offer attractive and reliable interactive capabilities, but their form-factor is often bulky, largely due to the need to integrate a sufficiently sized battery.

In this work, we explore how two common device form-factors---a ring and wristband---can work together to unlock a new design space of small, battery-free rings for subtle thumb-to-index input. 
Because today's rings for micro finger gestures must power sensing and communication modules, their usability is hindered by the need for a bulky battery~\cite{Kienzle2021ElectroRing, Nirjon2015TypingRing, Parizi2019AuraRing} or cumbersome power supply wiring~\cite{EFRing2023Chen, Liang2021Dual} in what would otherwise be a small, comfortable ring form factor.
On the other hand, many users already wear a wristband all day long for health and fitness applications and for their convenient visual display, but typical hand interactions provided by a wristband requires dynamic, expressive hand gestures~\cite{Jiang2022Survey, Laput2016ViBand, Yang2016Tomo2}, which can be fatiguing or impractical in daily use.
To fill this gap, we consider how a battery-free ring can complement a wristband by enabling a variety of reliable microgestures.

This paper introduces picoRing, a thumb-to-index finger input system enabling a battery-free, tiny, and lightweight ring-based input device~(\autoref{fig:overview}).
picoRing operates based on a pair of two coils: a ring-based fully-passive sensor coil and a wristband-based reader coil.
Using a coil-based readout technique called passive inductive telemetry~(PIT)~\cite{Takahashi2020TelemetRing,Takahashi2022Twin}, the wristband coil wirelessly monitors the passive response of the ring coil caused by user inputs.
First, the ring coil, which consists of a passive resonant coil connected to a physical switch, can passively convert the subtle thumb-to-index finger action (e.g., pressing the switch) into a unique passive response (i.e., the change of the ring's resonant frequency). 
Then, the wristband coil, which couples with the ring coil via a weak inductive field, can recognize the finger input by detecting the shift of the ring's resonant frequency.
Because the ring operates by modifying the inductive field in a passive manner and does not require active signal transmission, the ring can be battery-free.

Unlike semi-passive readouts like near-field communication~(NFC) and radio-frequency~(RF) backscatter~\cite{Saman2018video, Lin2020NFC}, fully-passive PIT-based systems~\cite{Takahashi2020TelemetRing, Takahashi2022Twin} can eliminate the need for powering up the ring, enabling a chipless and simple ring design. 
Prior PIT-based input systems~\cite{Takahashi2020TelemetRing} leveraged passive input for detecting taps on surfaces.
However, systems like TelemetRing~\cite{Takahashi2020TelemetRing}, which requires a large (\textit{e.g.}, \SI{9}{\cm}-diameter) uncomfortable wristband coil to capture the ring's weak response, is not suitable for wearable usage and is primarily designed for simple taps.
In contrast, picoRing stands out due to its versatility and practicality.
Specifically, picoRing introduces four types of rings of mass \numrange{1.5}{2.9}~\si{\g} for subtle thumb-to-index inputs including pressing, 1-D/2-D sliding, or scrolling with a network of different passive switches.
Moreover, picoRing increases the PIT sensitivity by about ten times through a design optimization process at a higher frequency band from \SI{13.5}{\MHz} to \SI{27}{\MHz}.
This results in a \SI{5}{\cm}-diameter-sized compact wristband coil demonstrating a stable readout distance of approximately \SI{13}{\cm}, regardless of finger bending or wrist in proximity with metallic items.
Such a compact implementation of the wristband and ring increases the practicality of this approach and allows integration of picoRing into standard ring and wristband accessories.

Our contribution is summarized as follows:
\begin{itemize}
    \item The practical design of a sensitive PIT technique to pair a battery-free ring coil with a compact wristband coil.
    \item The demonstration and technical evaluation of four types of rings, each below \SI{3}{\g} for subtle thumb-to-index input.
\end{itemize}

\section{Related work}

This section focuses on prior work which can recognize finger gestures by instrumenting an input system around the hand.
For more detail on other prior work, please refer to~\cite{Jiang2022Survey, Kienzle2021ElectroRing}.

\subsection{Instrumented hand}

A wristband serves as a convenient location for detecting finger movements based on force measurement~\cite{Artem2014WristFlex}, EM wave~\cite{Kim2022EtherPose}, inertial sensing~\cite{Laput2016ViBand}, capacitive coupling~\cite{Rekimoto2001GestureWrist}, electromyography~(EMG)~\cite{Saponas2009EMG},  computer vision~\cite{Wu2020Back}, etc.
Because the wristband infers finger gestures away from the wrist, the signal caused by subtle finger motions inevitably becomes small, challenging for the wristband to stably detect and accurately distinguish them~\cite{Wu2020Back, Chenhan2023aofinger}.
The use of machine/deep learning algorithm~\cite{Kim2022EtherPose} and sensor fusion approach~\cite{Chenhan2023aofinger} can improve the recognition accuracy of the various finger gestures, and also recognize user-defined input as customized input with a few-shot learning~\cite{Xu2022enabling}.
However, 1)  the approach is limited to within-user/short-term usage, requiring periodic calibration~\cite{Kim2022EtherPose}, 2) the wristband needs not subtle but relatively dynamic finger motions to accurately classify different gestures~\cite{Xu2022enabling}, or 3) the pure prediction accuracy significantly decreases when the wristband tries to recognize various subtle finger motions~\cite{Chenhan2023aofinger}. 
As a result, most wristband sensing rather focuses on the stable classification of expressive finger gestures~\cite{Jiang2022Survey, Laput2016ViBand}, possibly causing fatigue and discomfort for daily usage.
In contrast, a finger-mounted device like a ring offers the advantages of close-range sensing for finger gestures, enabling reliable detection of subtle finger movements using a simple sensor and algorithm~\cite{EFRing2023Chen, Gong2017Pyro, Kienzle2014LightRing, Meskers1998MagnetTracking, Nirjon2015TypingRing, Reyes2018SynchroWatch}.
Moreover, the sensing near the fingers can enable stable classification of micro, discrete finger gestures like tapping~\cite{Kienzle2021ElectroRing, Chenhan2023aofinger}, and mm-scale, continuous finger gestures such as sliding and scrolling~\cite{Ashbrook2011Nenya, EFRing2023Chen, Waghmare2023OptiRing}, allowing imperceptible interactions in everyday life.

\subsection{Ring-based input device}

A ring positioned close to the index finger is an ideal accessory for implementing a thumb-to-index finger input system. 
Furthermore, the ring design that leaves the fingertip uncovered is more acceptable for daily use compared to other finger-mounted form factors like nails~\cite{Chan2013FingerPad,Chen2016Finexus}.
Commercially, some smart rings (\textit{e.g.}, Oura Ring) are already available as healthcare monitoring.
To detect subtle finger inputs, many standalone rings have been developed~\cite{Fukumoto2003BodyCoupledFingerRing, Gong2017Pyro, Kienzle2014LightRing, Kienzle2021ElectroRing, Nirjon2015TypingRing}, but, prior rings are so bulky and heavy~($>\SI{10}{\g}$) due to the equipment of battery, signal processing circuit, and wireless communication module along with a sensor. 
Additionally, integrating these components into a small ring is complicated and costly.
In contrast, the pair of a ring and a wristband enables a slimmer ring design by relocating the signal processing module from the ring to the wristband~\cite{Parizi2019AuraRing, Takahashi2020TelemetRing}.
For example, a ring equipped with a passive magnet~\cite{Ashbrook2011Nenya} or an oscillating coil~\cite{Parizi2019AuraRing,Raab1979Magnetic}, which can transmit a varying signal related to finger movement to a signal-processing wristband, can be lightweight~($<\SI{5}{\g}$), compact, and tiny. 
Nevertheless, the ring itself remains bulky compared to the standard ring because it requires either a bulky magnet or a battery-driven oscillator to transmit the strong static or dynamic inductive signal to the wristband.

\subsection{Battery-free ring-based sensor}

To eliminate the battery from the ring connected to the wristband, RF backscatter~\cite{Saman2018video}, inductive power transfer~\cite{Takahashi2022WPT}, and PIT~\cite{Takahashi2020TelemetRing} are promising.
RF backscatter operates a low-powered semi-passive antenna tag by harvesting environmental energy and reflecting an incoming RF signal from a reader antenna.
Unfortunately, the EM interaction of the RF signal with the body can degrade communication performance, which requires a reader with \si{\W}-class high input power. 
Moreover, the signal path is affected by unrelated hand movements, which could be misclassified as false positive.
Inductive power transfer, which wirelessly transmits power to a ring-based coil via a wristband-based coil, also suffers from a low efficiency below $1\%$ due to a large size ratio of the tiny ring and the compact wristband and relatively long distance between the ring and the wristband~\cite{Takahashi2020TelemetRing}.
As a result, the input power from the wristband potentially exceeds \numrange{5}{10}~\si{\W} considering 1\% efficiency at best and over \SI{100}{\mW} power consumption of the communication module in the ring. 
Such high power levels are impractical for the power design of wristband devices.

In contrast, PIT based on an inductive link can reliably read out the sensor value from a fully-passive ring coil via a wristband coil without powering up the ring, as detailed in \S~\ref{sec:pit}.
However, the prior PIT proposed by TelemetRing~\cite{Takahashi2020TelemetRing}, which is most similar to picoRing's underlying technology, requires a \SI{9}{\cm}-diameter-sized large wristband coil due to its insufficient sensitivity.
This is not suitable for wearable applications as it leaves a big gap~(\SI{>3}{\cm}) around the wrist.
Unlike TelemetRing, picoRing offers distinct advantages including 1) the stable connection of the battery-free ring with a flexible \SI{5}{\cm}-diameter wrist-sized compact wristband, and 2) the sensing capability to detect various subtle finger movements using various battery-free rings.

\section{Theory of operation}

\begin{figure*}[t!]
  \centering
  \includegraphics[width=2.0\columnwidth]{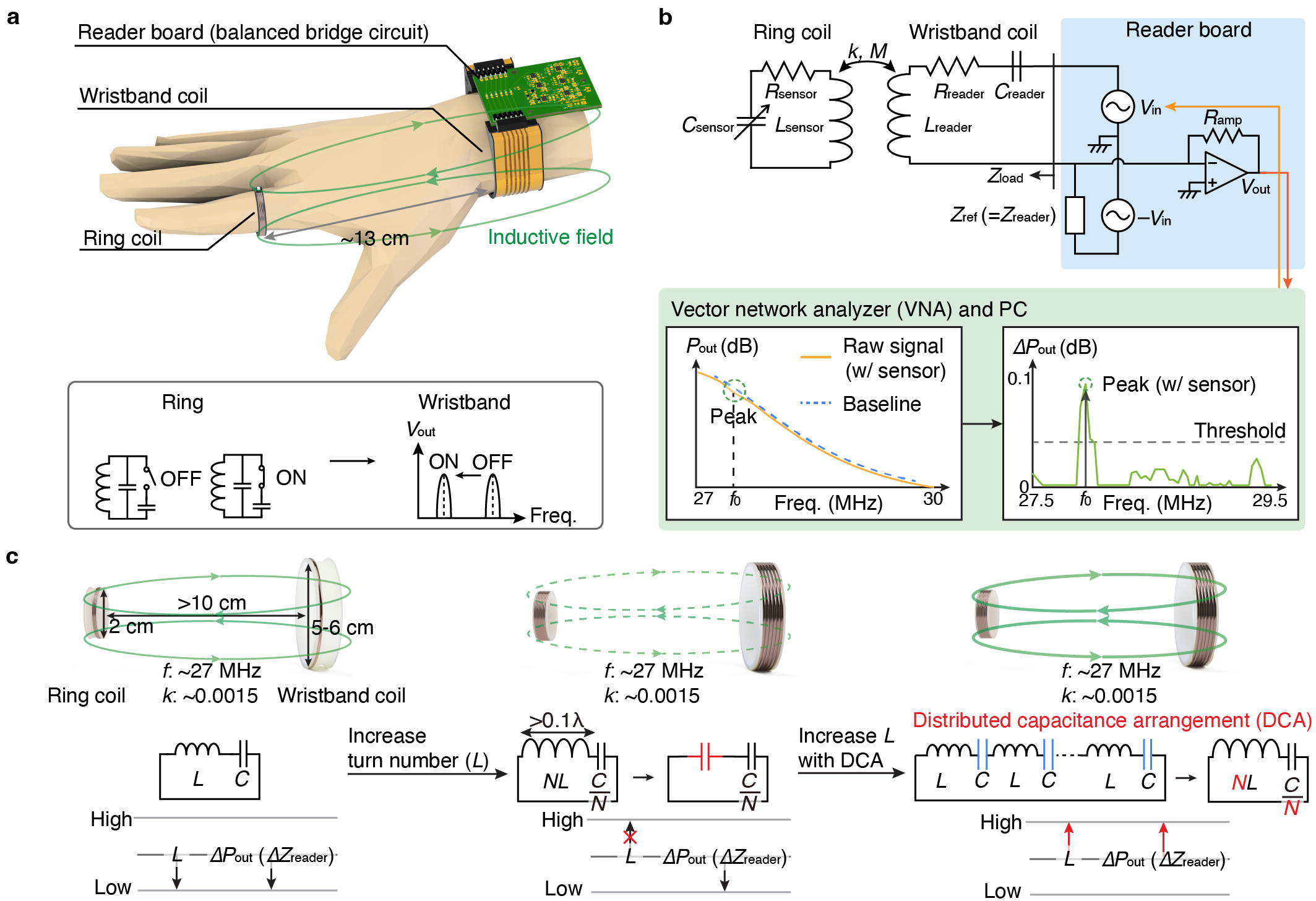}
  \caption{picoRing design. (a) The system illustration of picoRing. picoRing is based on passive inductive telemetry~(PIT), in which the wristband coil wirelessly monitors the passive response of the ring coil caused by user inputs. (b) The circuit diagram and signal processing of picoRing which consists of a ring-based fully-passive sensor coil and a wristband-based reader coil connected to a reader board. (c) Overview of distributed capacitance arrangement~(DCA) technique which can increase the passive response by enabling a high inductor at a high frequency.}
  \label{fig:design}
  \Description{The figure is an informational graphic depicting the design and functional principles of a wireless communication system involving a ring and wristband. Panel a illustrates the system on a mannequin hand, showing the placement of the ring and wristband coils and a reader board attached to the wristband, with indications of an inductive field and the direction of signal modulation from ON to OFF with changes in frequency. Below are circuit symbols representing the ring and wristband states. Panel b presents a detailed schematic of the system's electrical components and circuits, including mutual inductance between ring and wristband coils and reader board, along with a vector network analyzer (VNA) graph displaying the raw signal and threshold levels for signal output in decibels. Panel c shows two cylindrical visualization models depicting the spatial relationship between ring and wristband coils, accompanied by simplified circuit diagrams indicating strategies to modify the system's sensitivity by changing inductance and capacitance distribution, with corresponding effects shown on signal power differential graphs.}
\end{figure*}

\subsection{System overview}
\label{sec:system_overview}

picoRing consists of two main components:~1)~a ring-based fully-passive sensor coil~(i.e., ring coil) that passively changes its resonant frequency based on the user's thumb-to-index input to a passive switch such as a tactile or slider switch and 2)~a wristband-based reader coil~(i.e., wristband coil) that detects a peak in the frequency response corresponding to the resonant frequency of the ring coil~(\autoref{fig:design}a).
The working principle is as follows~(\autoref{fig:design}b):
First, the ring coil picks up the inductive field emitted by the wristband coil. 
Then, the ring coil causes a strong induced current in the wristband coil around the ring's resonant frequency based on Faraday's law of induction. 
Note that the resonant frequency of the ring is calculated as follows: $f_{0}=1/\left(2\pi\sqrt{L_{\rm sensor}C_{\rm sensor}}~\right)$ where $L_{\rm sensor}$ and $C_{\rm sensor}$ are inductance and capacitance of the ring coil.
The current also causes an impedance change in the wristband coil around $f_0$.
Finally, a switch-based variable capacitor connected to the ring coil can convert the user's input such as press to the shift of $f_{\rm sensor}$ by passively changing $C_{\rm sensor}$ via the on/off of the switch.
Finally, a reader board connected to the wristband coil can recognize a tiny impedance change in the frequency response using a sensitive impedance measurement circuit and a sensitive peak detection algorithm, as will be explained in \S~\ref{sec:pit} and \S~\ref{sec:signal_processing}.

\subsection{Sensitive passive inductive telemetry}
\label{sec:pit}

picoRing needs to increase the sensitivity of prior PIT techniques~\cite{Takahashi2020TelemetRing} because the large size ratio and long distance between the ring and wristband weaken the passive response significantly.
Basically, the passive response strengthens with the increase of the three factors: distance between wristband/ring coils~(coupling coefficient:~$k$), the inductance of the coils~($L$), and resonant frequency~($f_0$).
When the wristband coil inductively couples with the ring coil as shown in~\autoref{fig:design}c, the input impedance of the wristband coil~($Z_{\rm load}$) changes based on the passive response~($\Delta Z_{\rm reader}$) as follows:
\begin{align}
    Z_{\rm sensor} &= 
    \begin{cases}
    R_{\rm sensor} + j\left(\omega L_{\rm sensor} - \cfrac{1}{\omega C_{\rm sensor}}\right) & (f \ne f_{\rm 0})\\
    R_{\rm sensor} & (f = f_{\rm 0})
    \end{cases} \\
    Z_{\rm load} &= Z_{\rm reader} + \cfrac{(\omega M)^2}{Z_{\rm sensor}} \\
    &= Z_{\rm reader} + \Delta Z_{\rm reader} \label{eq:delta_load}
\end{align}
where $\omega~(=2\pi f)$ is an angular frequency, $M~(=k\sqrt{L_{\rm reader}L_{\rm sensor}})$ is mutual inductance between the wristband and ring coils, $Z_{\rm reader}$ and $Z_{\rm sensor}$ are impedance of the wristband and ring coils consisting of the coil loss~($R$), the coil inductance~($L$), and the chip capacitor for tuning a resonant frequency of the coil~($C$), respectively, and $\Delta Z_{\rm reader}$ is the impedance change caused by the ring coil.
$\Delta Z_{\rm reader}$ has a large peak around $f_{0}$ because $Z_{\rm sensor}$ remains only a small real impedance $R_{\rm sensor}$ at $f_0$ cancelling a large imaginary impedance $\omega L_{\rm sensor}$ by $1/(\omega C_{\rm sensor})$.
Thus, the wristband can estimate the resonant frequency of the ring by detecting the peak in $Z_{\rm load}$.
Unfortunately, the extremely weak $k$ below $0.002$ in picoRing makes $\Delta Z_{\rm reader}$ below \SI{1}{\mohm}, which makes it hard for the wristband to capture such a small peak.

To increase $\Delta Z_{\rm reader}$ under the low $k$, one simple solution is to design wristband and ring coils with a high $L$ or high $f_0$. 
However, the short wavelength at a high $f_0$ prevents increasing the turn number of the coil.
Thus, prior PIT uses either a high $L$~(>\SI{10}{\uH}) at a low $f_0$~(\si{\kHz}) or a low $L$~(<\SI{1}{\uH}) at a high $f_0$~(tens of \si{\MHz}), which inevitably results in small $\Delta Z_{\rm reader}$ under the low $k$.
To solve this trade-off, picoRing employs distributed capacitance arrangement~(DCA) technique, enabling a high $L$ at a high $f_0$~\cite{Cook1982ANMR}.
DCA separates a long coil into multiple shorter coils by inserting multiple chip capacitors in series.
With this configuration, each short coil can behave as a small $L$~(\autoref{fig:design}c).
Because the multiple small $L$ are connected in series, the coil with a large turn number comprised of a long wire can behave as a high $L$~(\si{\uH}) at the high $f_0$~(\si{\MHz}), which can increase $\Delta Z_{\rm reader}$ under the low $k$.

Although DCA strengthen the passive response, it is still relatively weak~($\Delta Z_{\rm reader} \approx \SI{50}{\mohm}$) because of the low $k$~($\sim0.001$).
To detect such a small impedance change, picoRing then employed a balanced bridge circuit~\cite{Takahashi2020TelemetRing}.
Unlike a standard impedance measurement circuit which measures the target impedance itself, the bridge circuit can measure only the impedance change by using a differential structure~(\autoref{fig:design}c).
Thus, the bridge circuit is useful for accurately measuring a small impedance change in an otherwise high impedance like with a thermistor or strain gauge.
Here, picoRing uses the bridge circuit simply by matching a reference load~($Z_{\rm ref}$) composed of chip elements with the wristband coil~(\textit{i.e.}, $Z_{\rm reader} = Z_{\rm ref}$).
With this, the voltage output of the bridge circuit $V_{\rm out}$ can be calculated as follows with Eqn.~\ref{eq:delta_load}, $Z_{\rm ref} = Z_{\rm reader}$, and $Z_{\rm reader} \gg \Delta Z_{\rm reader}$:
\begin{align}
    V_{\rm out} &=  -R_{\rm amp} \left(\cfrac{V_{\rm in}}{Z_{\rm load}} - \cfrac{V_{\rm in}}{Z_{\rm ref}}\right)\\
    &\approx  
    \begin{cases}
         &  R_{\rm amp} \cfrac{\Delta Z_{\rm reader}}{Z_{\rm ref}^2}V_{\rm in} ~(\mbox{w/ ring coil})\\
         & 0 ~(\mbox{w/o ring coil i.e.,}~\Delta Z_{\rm reader}=0)
    \end{cases} \label{eq:output_bridge}
\end{align}
where $V_{\rm in}$ is the input voltage of the bridge circuit, $R_{\rm amp}$ is the amplifier factor.
Note that the real part of $Z_{\rm reader}$ is adjusted over \SI{50}{\ohm} with a chip resistor to meet the relationship of $Z_{\rm reader} \gg \Delta Z_{\rm reader}$.
Eqn.~\ref{eq:output_bridge} indicates that $V_{\rm out}$ is sensitive to $\Delta Z_{\rm reader}$ by the matching process.
Thus, the small peak~($\Delta V_{\rm out}$) appears in $V_{\rm out}$ via the bridge circuit~(\autoref{fig:design}c).

\subsection{Resonance detection}
\label{sec:signal_processing}

\begin{figure*}[t!]
  \centering
  \includegraphics[width=2.0\columnwidth]{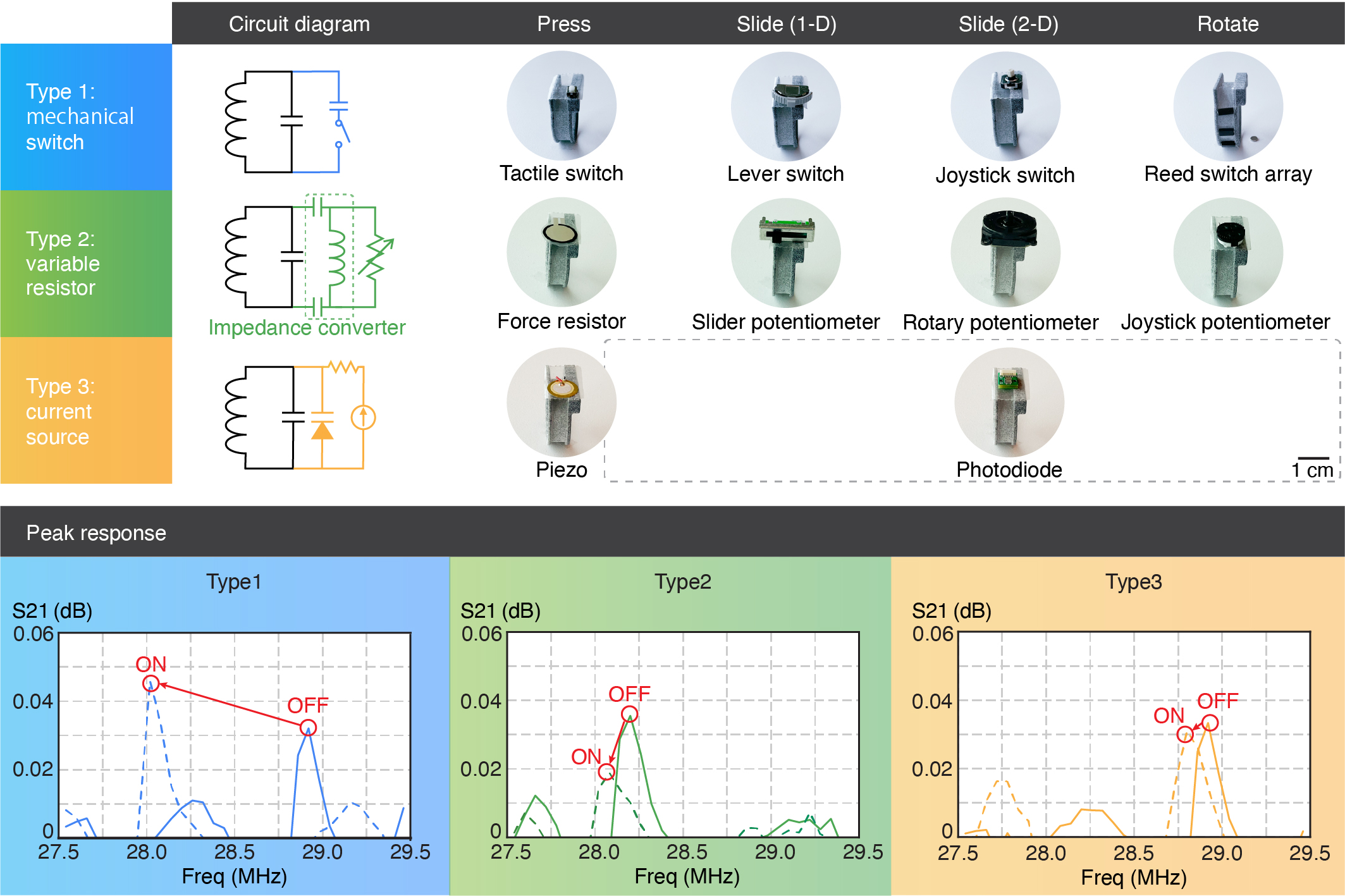}
  \caption{Design of passive variable capacitor. Among three types, picoRing uses type 1.}
  \label{fig:sensor}
  \Description{The image is a comprehensive illustration of different types of input mechanisms for electronic devices, categorized by circuit diagram, peak response graph, and the physical form of the inputs. There are three main categories: Type 1 mechanical switch, Type 2 variable resistor, and Type 3 current source, each with an associated circuit diagram. The peak response graphs show the voltage output as a function of frequency with notable peaks at sensor frequencies. The input mechanisms include tactile switch, lever switch, joystick switch, and a reed switch array for pressing actions; force resistor, slider potentiometer, and rotary potentiometer for sliding actions in 1-D or 2-D movements; and finally, piezo and photodiode for rotational interactions. Each input device is shown as a picture of the actual component, with a scale for size reference at the bottom right.}
\end{figure*}

Finally, picoRing needs to stably recognize the small peak~($\Delta V_{\rm out}$) in $V_{\rm out}$ despite the variations of $V_{\rm out}$ owing to the capacitive coupling with the body, the eddy current caused by the metal, EM noise from electric appliances, etc.
To solve these challenges, picoRing uses a sensitive peak detection algorithm, which can find the peak through the error caused by a least-squared fitting of the frequency response~(\autoref{fig:design}c). 
First, the least-squared fitting function~(\textit{e.g.}, \textit{scipy.polyfit}) estimates a smooth baseline from the raw magnitude output ($P_{\rm out}=20\log_{\rm 10} V_{\rm out}$) from the bridge circuit.
Notably, the baseline is almost similar to the average line of $P_{\rm out}$, which removes a small peak ($\Delta P_{\rm out}$) and noises in $P_{\rm out}$.
Then, the fitting error containing $\Delta P_{\rm out}$ is derived by calculating the difference between the baseline and the $P_{\rm out}$. 
Finally, a standard peak detection algorithm~(\textit{e.g.}, \textit{scipy.find\_peaks}) can find a clear peak against the small noise~(\autoref{fig:design}c). 
The peak threshold was set to \SI{0.02}{\dB}, considering the dB RMS of a vector network analyzer~(VNA)~(\SI{0.002}{\dB}), which converts $V_{\rm out}$ into $P_{\rm out}$.
Because the baseline is updated in real-time, the peak detection can be robust against the amplitude and frequency drift of $V_{\rm out}$.

Unlike the prior PIT using DCA around \SI{13.5}{\MHz} frequency band~\cite{Takahashi2020TelemetRing}, we note that picoRing uses DCA around the two times higher frequency band around \SI{27}{\MHz} to further increase prior PIT sensitivity, in addition to the unique peak detection algorithm.
As a result, picoRing achieves sensitive readout of the fully-passive ring coil via the weakly-coupled compact wristband coil.
Mathematically, the most critical technique to increase the SNR, or $P_{\rm out}$ is the distributed capacitance technique. 
The distributed capacitor increases the $P_{\rm out}$ by approximately \SI{39}{\dB}, while the bridge increases the $P_{\rm out}$ by approximately \numrange{19}{25}~\si{\dB} at \SI{29}{\MHz}. 
The peak detection does not increase $P_{\rm out}$, but it can detect a small peak.
We experimentally checked that our peak detection can stably detect a small peak over \SI{0.1}{\dB} against the output drift (see \S~\ref{sec:snr_for_EMnoise}), resulting in the about $10x$ higher SNR.

\section{Design of battery-free ring coil}
\label{sec:ring_design}

In the context of AR/VR and human-computer interactions, subtle thumb-to-index finger inputs are frequently used as always-available and easy-to-use wearable controllers~\cite{Jiang2022Survey}.  
For example, press, slide, and scroll finger inputs can complement touch or voice controls within the existing wearable devices to quickly access features, make selections, or control displays.
Thus, this section explores how to recognize the press, slide, or scroll finger inputs via the ring coil.
Because the wristband coil estimates the sensor state of the ring coil through the shift of the ring's resonant frequency~($f_0=1/(2\pi\sqrt{L_{\rm sensor}C_{\rm sensor}})$), picoRing focuses on how to design a passive variable capacitor reacting to the above finger inputs~(\autoref{fig:sensor}).
One could also create a variable inductor, but this is more complicated, so in this work we limited our design space to variable capacitors.

There are three types of passive variable capacitors: type 1 based on a mechanical switch, type 2 based on a set of a variable resistor and an impedance converter circuit, and type 3 based on a set of a varactor and a passive current source~(piezo or photodiode).
Among the three types, picoRing needs a small and passive variable capacitor to embed the capacitor in the tiny and battery-free ring.
Type 1 has low-cost and size advantages in addition to requiring users to only connect a tiny chip capacitor to the switch.
As for type 2, the commercially-available variable resistor like the potentiometer is too bulky for a ring.
Similar to type 2, type 3 requires a bulky battery because the passive current from the piezo or photodiode is too weak without a signal amplifier.
Moreover, the frequency shift of type 2 and type 3 is much smaller than the one of type 1.
Thus, picoRing adopts type 1 sensors to a ring coil, as described in~\S~\ref{sec:impl_ring}.
Specifically, picoRing uses a tactile switch, lever switch, or joystick reacting to press or 1-D/2-D slide input, respectively.
Furthermore, picoRing uses the combination of the multiple reed switches and magnets to detect the scroll of the ring.
Note that the use of the multi-functional switch could combine the above four switches together, but, currently-available multi-functional switch~(\textit{e.g.}, EVQWJN005, Panasonic) needs battery-driven active sensors near the switch to distinguish the direction of sliding or scrolling. 
Thus, this paper uses different passive mechanical switches for detecting pressing, sliding, or scrolling interactions.

\section{Implementation}
The following sections describe the implementation of both the wristband reader and ring-based sensor coil.

\subsection{Wristband-based reader coil}
\label{sec:impl_wrist}

\begin{figure}[t!]
  \centering
  \includegraphics[width=1.0\columnwidth]{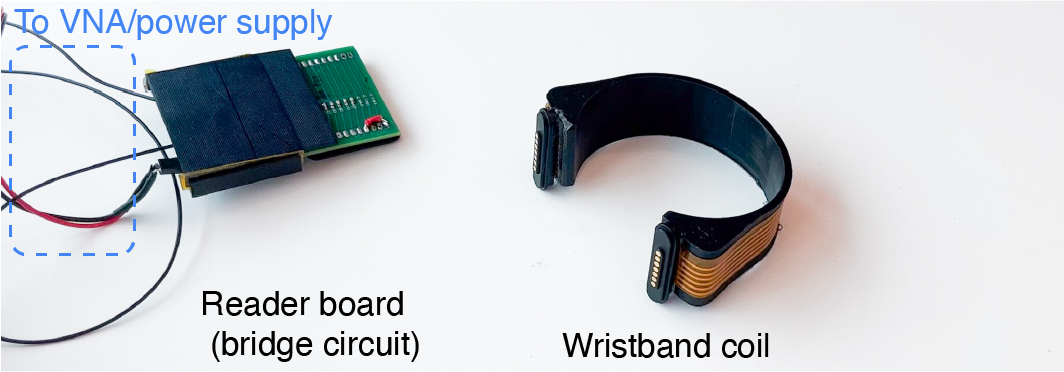}
  \caption{Implementation of a wristband coil.}
  \label{fig:impl}
  \Description{The image shows two key components of a wearable technology system on a white background. On the left side, there is a compact electronic device labeled as "Reader board (bridge circuit)" with wires leading off the image indicating a connection to a Vector Network Analyzer (VNA) or power supply. On the right side, there is a "Wristband coil" that appears to be a flexible band with a coil pattern visible on the surface, suggesting its function as part of a wireless communication or sensing system. The reader board is mounted on fabric and connected to the wristband via wires, illustrating an open setup presumably for development or testing purposes.}
\end{figure}

The wristband coil consists of 1) a resonant coil implemented by flexible PCBs, 2)~a flexible wristband implemented by 3D printing of elastomeric polyurethane, and 3)~a readout board containing two magnetic connectors, the balanced bridge circuit connected to the external laptop-sized VNA~(PicoVNA 106, Pico Technologies) and a laptop~(\autoref{fig:impl}).
To construct a proof-of-concept prototype, we selected the bulky PicoVNA for fast signal emission and acquisition, but, a set of a tiny VNA chip~(ADL5960, Analog Devices) and a tiny FPGA~(ICE40LP8K, Lattice Semiconductor) could be useful to implement a compact, low-powered, self-contained readout board.
The turn number, size, and thickness of the coil are $6$, \numproduct{4.7 x 6.0}~\si{\cm}, and \SI{2}{\mm}, respectively.
In addition, the resonant frequency, inductance, and resistance of the coil are \SI{27}{\MHz}, \SI{3.7}{\uH}, and \SI{55}{\ohm}, respectively. 
Using DCA, the coil was tuned by soldering eighteen \SI{170}{\pF} chip capacitors and one \SI{51}{\ohm} chip resistor in series.
The coil with the larger turn number can further increase $\Delta Z_{\rm reader}$, but, accurate impedance matching with the chip elements becomes harder. 
Thus, the turn number of the coil was set to $6$.

The coil was snapped on the readout board via magnetic spring connectors~(Magnetic Connector $6$ Contact Pins, Adafruit Industries).
The bridge circuit follow a similar implementation to \cite{Takahashi2020TelemetRing} and was implemented by multiple amplifiers available for a wide frequency band~(LTC6228, Analog Device) and connected to a \SI{5}{\V} portable charger (Portable Charger, Anker).
In total, the bridge consumed \SI{430}{\mW}~($=\SI{5}{\V}\times\SI{86}{\mA}$). 
Note that the power consumption could be significantly lowered~($\approx \SI{20}{\mW}$) by simply replacing the current amplifier with the low-powered amplifier~(e.g., LTC6253, Analog Device Inc.).
$R_{\rm amp}$ was set to \SI{100}{\ohm} considering the gain product of the amplifier.
The reference load was implemented by the chip elements.
The impedance error between the reference load~($Z_{\rm ref}$) and the coil~($Z_{\rm reader}$) was approximately \numrange{5}{10}\%, which was sensitive enough to recognize the small $\Delta Z_{\rm reader}$.
With this setup, the VNA connected to the bridge circuit inputs a sweep signal to the bridge circuit ranging from \SI{27}{\MHz} to \SI{30}{\MHz} in steps of \SI{60}{\kHz} with a bandwidth of \SI{10}{\kHz} and an input power of \SI{1}{\mW}.
Then, the VNA acquires the frequency response from the bridge circuit at approximately \SI{5}{fps} and the laptop analyzes the response to detect the resonant frequency of the ring coil.
Note the input power itself is a less crucial parameter since picoRing is required to measure the coupled ring's impedance through the power ratio of the input power and the output power in the wristband coil.
For instance, even when we increased the input power above \SI{1}{\mW}, the SNR of the picoRing remained almost unchanged.

\subsection{Ring-based sensor coil}
\label{sec:impl_ring}

The ring consists of a coil comprised of $24$ AWG copper wire, a ring base, and a rigid sensor board.
The turn number and size of the ring coil are \numrange{7}{8}, \SI{1.9}{\cm} diameter with the width of \numrange{4}{9}~\si{\mm}.
Note that the turn number was determined based on \S~\ref{sec:eva_snr_for_turn_number} evaluation.
Moreover, our preliminary investigations revealed that ISM frequency band above \SI{40}{\MHz} leads to inevitable electromagnetic interactions and increased impedance in the ring coil when worn on the hand, resulting in an SNR decrease. 
Specifically, we found that ring coils resonating at \SI{13.56}{\MHz} or \SI{27.12}{\MHz} exhibit minimal impedance change (5-10\%), whereas coils tuned to \SI{40.68}{\MHz} experience significant impedance changes (over 50\%) when worn. 
Consequently, we selected the \SI{27.12}{\MHz} band as the ring's resonant frequency.
The ring base was SLS-printed with Nylon material.
The sensor board was designed based on the circuit diagram of type1~(\autoref{fig:sensor}), which consists of a pair of a mechanical switch and chip capacitors.
Four types of mechanical switches were employed to detect basic subtle one-handed inputs:~a tactile switch for pressing input, a lever switch and a joystick switch for sliding input in 1-D and 2-D, and a set of reed switches for scrolling input. For more detail, please refer to \S~\ref{sec:example} and \autoref{tab:ring}.

\begin{table}[t!]
    \centering
    \caption{Physical specification of picoRing prototypes.}
    \begin{tabular}{llll}\toprule
        \textbf{Ring} & \textbf{Weight} & \textbf{Width} & \textbf{Thickness}\\\hline
        picoRing \textit{press} & \SI{1.5}{\gram} & \SI{4.2}{\mm} &  \SI{1.5}{\mm} \\
        picoRing \textit{slide} & \SI{2.9}{\gram} & \SI{5.6}{\mm} & \SI{1.5}{\mm} \\
        picoRing \textit{joystick} & \SI{2.7}{\gram} & \SI{5.6}{\mm} & \SI{1.5}{\mm}\\
        picoRing \textit{scroll} & \SI{2.5}{\gram} & \SI{8.8}{\mm}& \SI{3.2}{\mm} \\
        \bottomrule
    \end{tabular}
    \Description{The table displays the physical specifications of four picoRing prototype models. The table lists each ring model with its corresponding weight, width, and thickness. The picoRing press weighs 1.5 grams, has a width of 4.2 mm, and a thickness of 1.5 mm. The picoRing slide weighs 2.9 grams and has both a width and a thickness of 5.6 mm and 1.5 mm respectively. The picoRing joystick has a similar weight of 2.7 grams, a width of 5.6 mm, and a thickness of 1.5 mm. Lastly, the picoRing scroll weighs 2.5 grams, is 8.8 mm wide, and 3.2 mm thick.}
    \label{tab:ring}
\end{table}

\section{Technical evaluation}

To characterize how the peak signal varies with the coil parameters, finger movements, or surrounding objects, this section evaluates the signal-to-noise ratio~(SNR) of picoRing for various conditions.
To decouple the readout board from the common ground, we inserted two RF transformers~(T1-1-X65+, Mini Circuits).
Because of the inductive coupling's resilience against the body, the SNR characteristics of picoRing shows minimal variations for similar hand sizes. 
Thus, we conducted the following evaluations for a single user~(man, 20s).

\subsection{SNR for turn number}
\label{sec:eva_snr_for_turn_number}

First, we evaluated the SNR for the turn number of the ring coil to decide the proper turn number of the ring coil.
The ring coil was tuned at \SI{29}{\MHz}, and the turn number ranged from $3$ to $9$ in the step of $1$.
The distance between the ring and the wristband was set to \SI{13}{\cm}.
The SNR was calculated as follows using S21 $100$ outputs~$\left(P_{\rm out}~(\si{\dB}) = 20\log_{\rm 10} V_{\rm out} \right)$ of the VNA:
\begin{align*}
    \mbox{SNR} = \cfrac{\mbox{mean}~\left(P_{\rm out~w/~sensor}\right) - \mbox{mean}~\left(P_{\rm out~w/o~sensor}\right)}{\mbox{std}~\left(P_{\rm out~w/o~sensor}\right)} \label{eq:snr}
\end{align*}
SNR for the turn number was described in \autoref{tab:eva_snr_for_number}.
The result indicated that the higher turn number results in a higher SNR, although the SNR varies little above $7$ turns.
This is because the coil loss~($R_{\rm sensor}$) owing to the long wire and proximity effect between the wires becomes non-negligible as the turn number increases.
Therefore, this paper employed $7$ or $8$ as the turn number of the ring coil.
Following the next sections, we measured the SNR of the ring coil with turn number of $8$.

\begin{table}[t!]
    \centering
    \caption{SNR, $L_{\rm sensor}$, $R_{\rm sensor}$, and the capacitor number for the turn number of a sensor coil tuned around $\mathbf{29}$~MHz.}
    \begin{tabular}{llllllll}\toprule
        \textbf{Turn number} & $\mathbf{3}$ & $\mathbf{4}$ & $\mathbf{5}$ & $\mathbf{6}$ & $\mathbf{7}$ & $\mathbf{8}$ & $\mathbf{9}$\\\hline
        \rowcolor{gray!20}
        SNR & $6.2$ & $8.6$ &$11.6$ &$13.5$ &$17.0$ &$18.2$ &$17.5$  \\
        \textbf{$L_{\rm sensor}$ (\si{\uH})} & \num{0.34} &  \num{0.56} & \num{0.85}& \num{1.2} & \num{1.4}& \num{1.8}& \num{2.1} \\
        \textbf{$R_{\rm sensor}$ (\si{\ohm})} & $0.89$ & $1.1$ & $1.5$ & $1.8$ & $2.0$ & $2.6$& $3.4$ \\
        Capacitor number & $1$ & $2$ & $2$ & $2$ & $3$& $3$& $3$ \\
        \bottomrule
    \end{tabular}
    \Description{The table lists the signal-to-noise ratio (SNR), inductance (L_sensor), resistance (R_sensor), and the number of capacitors for sensor coils with turn numbers ranging from 3 to 9, all tuned around 29 MHz frequency. As the turn number increases from 3 to 9, SNR increases from 6.2 to 17.5. The inductance values range from 0.34 µH to 2.1 µH, resistance values range from 0.89 Ω to 3.4 Ω, and the number of capacitors ranges from 1 to 3 depending on the coil's turn number.}
    \label{tab:eva_snr_for_number}
\end{table}

\begin{figure*}[t!]
  \centering
  \includegraphics[width=2.0\columnwidth]{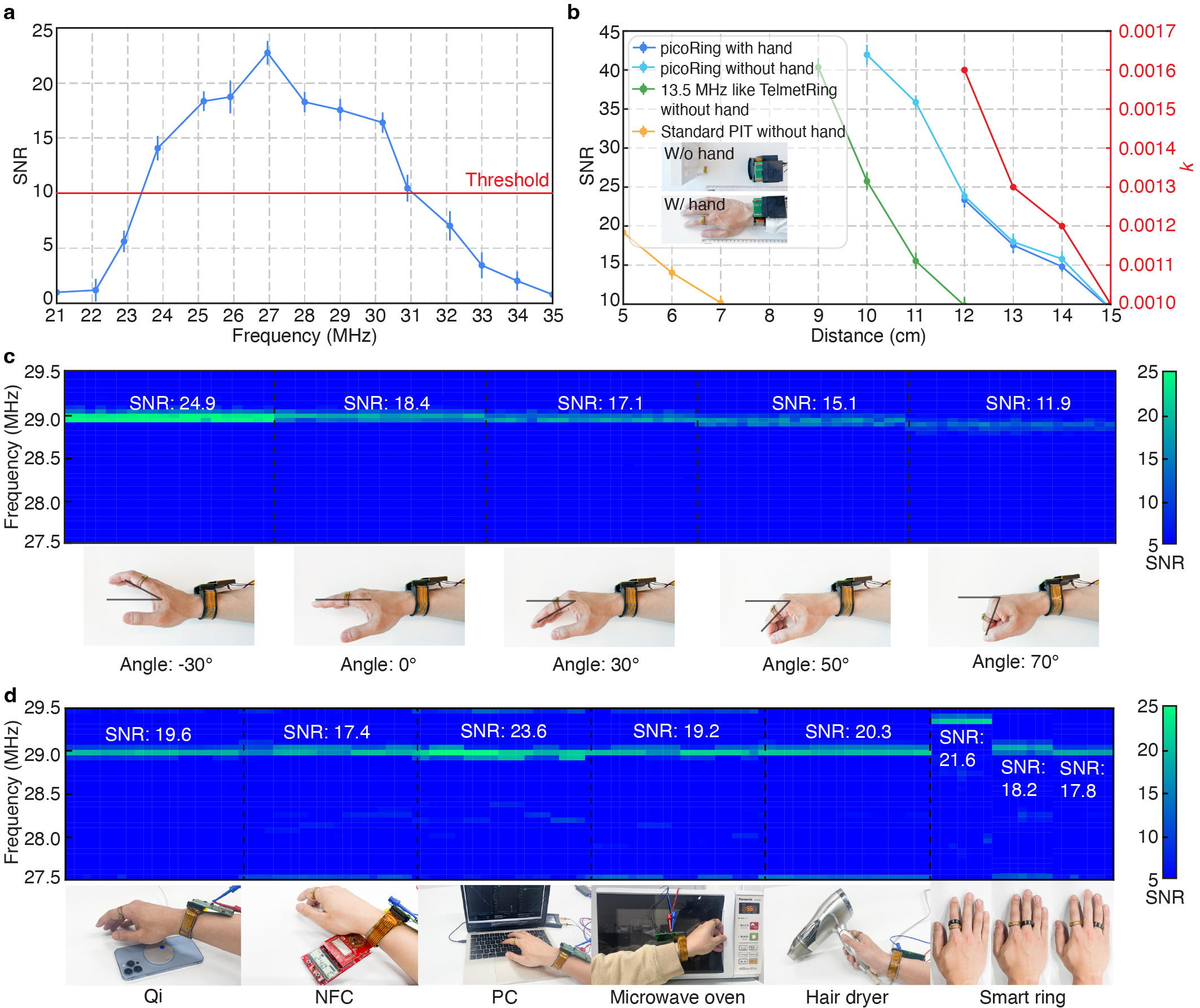}
  \caption{SNR evaluation of picoRing. (a) SNR for the resonant frequency of the ring coil. (b) SNR for the distance between the wristband and the ring. (c) SNR for different finger bending angles (d) SNR for proximity metallic items.}
  \label{fig:evaluation}
  \Description{The image is a compilation of charts and photographs illustrating a scientific study on signal quality for a smart ring device across various conditions. Panel a features a line graph plotting Signal-to-Noise Ratio (SNR) against frequency with a peak performance highlighted and a threshold line drawn in red. Panel b shows another line graph comparing SNR values for different distances with labels indicating measurements with hand and without hand, and has inset images of the device on a hand and standalone. Panel c and d contain a series of spectrograms with color-coded SNR values and corresponding poses or situations, demonstrating the performance of the device during different hand angles and when interacting with various electronic devices or appliances such as a Qi charging pad, Near-Field Communication (NFC), personal computer (PC), microwave oven, hair dryer, and a comparison with another smart ring. Insets show a hand wearing the device in specified angles and in use near different appliances to contextualize the spectrogram data.}
\end{figure*}

\subsection{SNR for frequency}
\label{sec:snr_for_freq}

To check the acceptable resonant frequency of the ring, we then examined the SNR for resonant frequencies ranging from \SI{20}{\MHz} to \SI{40}{\MHz} in \SI{1}{\MHz} step~(\autoref{fig:evaluation}a).
The distance between the ring and the wristband was set to \SI{13}{\cm}.
The result showed that the sensitive readout band~($\mbox{SNR}>10$) was \numrange{24}{31}~\si{\MHz} because the impedance matching of the bridge circuit was adjusted at the center of \SI{27}{\MHz}.
Note that at SNR below $10$, or $\Delta P_{\rm out}$ below $\SI{0.02}{\dB}$, it was difficult to distinguish the noise's peak robustly, as detailed in \S~\ref{sec:input_accuracy}.
Therefore, we reject false positives by setting the peak threshold to be $10x$ higher than the noise floor. 
Based on the result, we used \numrange{27.5}{29.5}~\si{\MHz}, which is a large enough band to monitor the peak shift of various picoRings, as will be described in \S~\ref{sec:example}.

\subsection{SNR for distance}
\label{sec:snr_for_distance}

Next, we evaluated the SNR for the readout distance to confirm the available distance between the ring and wristband coils.
\autoref{fig:evaluation}b illustrates the result of SNR.
First, we measured the SNR with and without the hand and confirmed that the SNR curve was almost the same, which means the inductive coupling between the coils interacts little with the body.
Then, we compared picoRing (with no hand) to two baselines: (i) a standard PIT coil without DCA or a bridge circuit, and (ii) a PIT coil tuned at \SI{13.5}{\MHz} similar to TelemetRing~\cite{Takahashi2020TelemetRing}.

In (i), the wristband coil was a $1$-turn, \SI{50}{\ohm} coil, and the ring coil was a $3$-turn resonant coil tuned at \SI{29}{\MHz} and the VNA measured S11~(impedance) of wristband coil directly without the bridge circuit.
In (ii), the wristband coil was a $6$-turn resonant coil tuned at \SI{13.5}{\MHz} and the ring coil was an $8$-turn resonant coil tuned at \SI{15}{\MHz}.
Because picoRing uses a two times higher frequency than (ii), the $P_{\rm out}$ can increase at least $12 P_{\rm out}(\approx 20\log_{\rm 10} 2^2 V_{\rm out})$ times compared to (ii).
The result shows that picoRing using the higher resonant frequency around \SI{27}{\MHz} increases its SNR by approximately $13$ compared to (i)(ii) and the available readout distance of picoRing~($\mbox{SNR}\geq 10$) reaches up to \SI{15}{\cm}~($k\geq 0.001$) with no misalignment between the coils.
Considering the SNR decrease owing to the finger angle~(\S~\ref{sec:snr_for_finger}), the available readout distance of the current picoRing is \SI{13}{\cm}~($x2.3$ reader's diameter).

\subsection{SNR for finger bending}
\label{sec:snr_for_finger}

\begin{figure*}[t!]
  \centering
  \includegraphics[width=2.0\columnwidth]{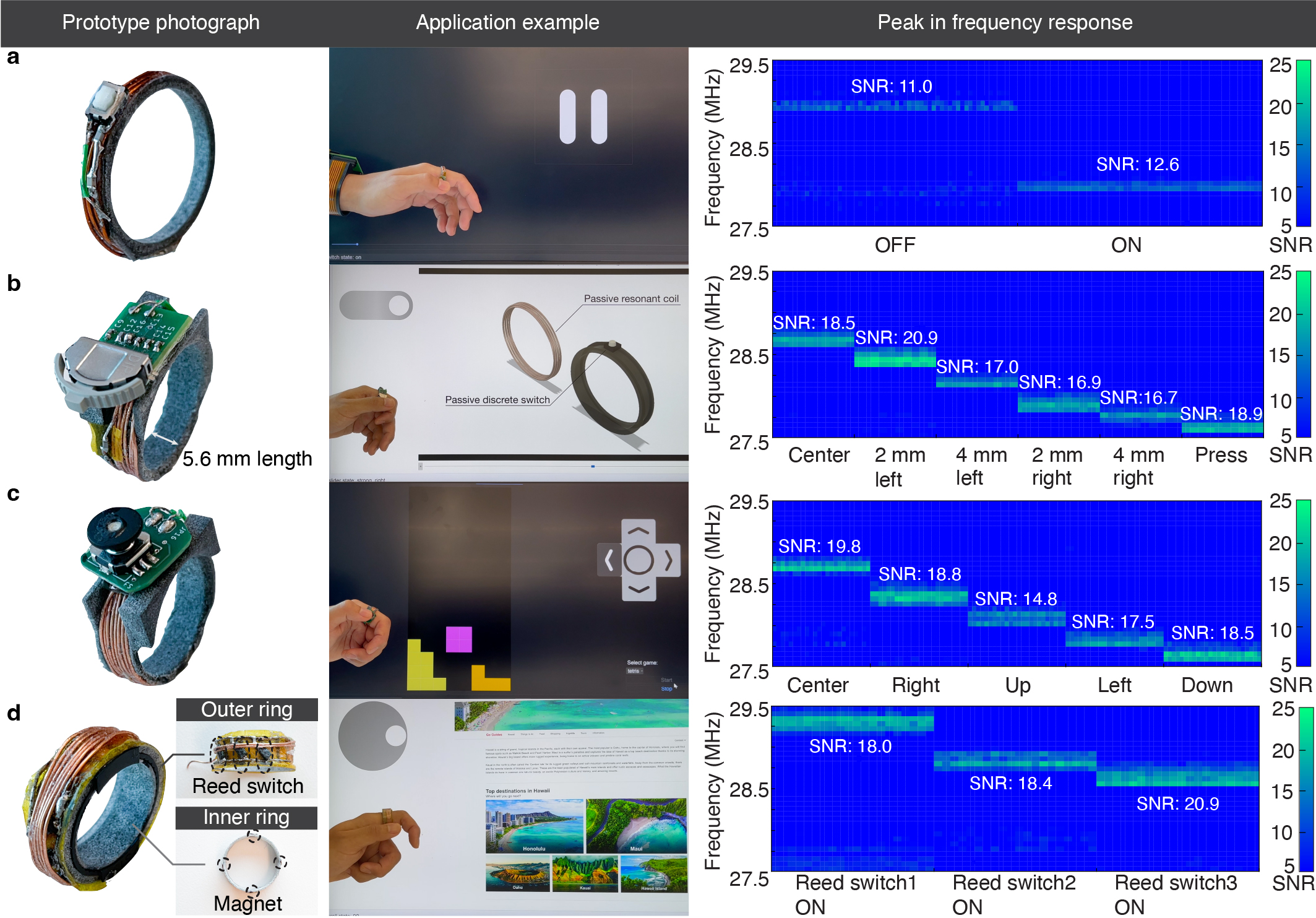}
  \caption{picoRing demonstration. Photograph, application example, and peak in the frequency response of picoRing (a) \textit{press}, (b) \textit{slide}, (c) \textit{joystick}, and (d) \textit{scroll}.}
  \label{fig:application}
  \Description{The figure displays various components and usage scenarios of a wearable smart device technology. On the left, prototype photographs are provided for different iterations of the device, including a ring (a), a wristband with mounted electronics (b), an electronic board placed on a fabric wristband (c), and a wristband with integrated magnetic switch technology (d). In the middle column, application examples are shown where a hand interacts with a digital interface, demonstrating touchless gesture commands like swipe and press. The rightmost column presents spectrograms labeled with Signal-to-Noise Ratio (SNR) values, illustrating the frequency response peaks of the device when performing various gestures such as pressing, moving fingers in different directions, and activating reed switches. Each spectrogram provides a visual representation of the effect these interactions have on the signal frequency and strength, correlating with the SNR readings.}
\end{figure*}

Then, we examined the SNR for finger angle because the inductive coupling~($k$) decreases due to the coil misalignment.
\autoref{fig:evaluation}c shows the time-series SNR spectrum and the average SNR against the five distinct finger angles~(\SI{-30}{\degree}, \SI{0}{\degree}, \SI{30}{\degree}, \SI{50}{\degree}, or \SI{70}{\degree}).
The distance between the ring and the wristband was set to \SI{13}{\cm}, and the ring was tuned at \SI{29}{\MHz}.
The result shows that picoRing can successfully capture the ring's peak at \SI{29}{\MHz} against up to \SI{70}{\degree} finger bending, which is sufficient for the daily hand movement.

\subsection{SNR for proximity metal}
\label{sec:snr_for_EMnoise}

Lastly, we measured the SNR against six kinds of metallic objects in the proximity of the hand because the eddy current or EM noise generated from metallic electrical appliances could inductively influence both the ring and wristband coil.
\autoref{fig:evaluation}d shows the time-series SNR spectrum and average SNR for the six items including Qi wireless charger~(Magsafe Charger, Apple) attached to a smartphone~(iPhone 13 Pro Max, Apple), a set of an NFC reader~(DLP-7970ABP, Texas Instruments) and an NFC tag~(TIDM-RF430-TEMPSENSE, Texas Instruments), a laptop~(MacBook Air M2, Apple), a microwave oven~(NE-EH228, Panasonic), a hair dryer~(KHD-9300, KOIZUMI), and a smart ring~(SOXAI RING 1, SOXAI).
Note that the SNR evaluation was conducted during operation of the appliances such as \SI{15}{\W} charging of the smartphone with the Qi charger, \SI{200}{\mW} charging and communication of the NFC tag with the NFC reader, \SI{700}{\W} heating of the rice with the oven, etc.
The result indicates that picoRing successfully detects a sharp peak at \SI{29}{\MHz} except for the smart ring placed near the ring coil.
This is mainly because 1)~for the readout board, the passive response from the ring coil is much higher than EM noise, and 2)~the peak detection algorithm is robust against the output drift caused by the eddy current, as explained in~\S~\ref{sec:signal_processing}.
Unfortunately, the resonant frequency of the ring coil is shifted up when the  metallic ring within \SI{1}{\cm} distance of the ring coil~(\textit{e.g.,} the metallic ring is worn either index or middle fingers) constructs a strong inductive coupling with the ring coil.
Future work could utilize such a peak shift up as the hint of metal detection near the ring.

\section{picoRing examples}
\label{sec:example}

This section presents four examples of picoRing~(\autoref{fig:application}).
Because of its easy-to-carry and always-available advantages, picoRing can potentially serve as a versatile input controller for ubiquitous devices, users, and situations.
Although the current prototype is divided into four input structures and is less adaptive, future iterations could feature a more flexible and integrated design to accommodate a wider range of interactions, as detailed in \S~\ref{sec:limit}.

\subsection{picoRing \textit{press}}
\label{sec:app_press}

First, \autoref{fig:application}a shows picoRing \textit{press} for a music controller.
picoRing \textit{press} consists of $7$-turn $2$-layer resonant coil with a tiny tactile switch~(EVQ-P2P02M, Panasonic), and the resonant frequency in \textit{on/off} state was tuned at \num{28.9} and \SI{28.0}{\MHz}, respectively.
Although the thumb-to-index press action is the one of most fundamental inputs in wearable computing, the prior rings are bulky and heavy due to the battery equipment. 
Unlike them, picoRing \textit{press} is tiny and lightweight~(\SI{1.5}{\gram}) because picoRing can replace a bulky battery and a sensing/communication module with a lightweight fully-passive coil~(\autoref{tab:ring}).
Thus, the battery-free, lightweight implementation of picoRing \textit{press} can provide continuous, natural interactions without worrying about the battery management.

\subsection{picoRing \textit{slide} or \textit{joystick}}
\label{sec:app_slide}

Then, \autoref{fig:application}b-c show picoRing \textit{slide} designed for a movie controller and \textit{joystick} for a game controller, respectively.
picoRing \textit{slide} that can detect center~(idle), \SI{2}{\mm} or \SI{4}{\mm} left, \SI{2}{\mm} or \SI{4}{\mm} right slide, and press input, consists of a ring base and an $8$-turn resonant coil with a lever switch~(PLMG5-GH-V-T/R, Diptronics).
The resonant frequency of five types of slide and press states was tuned at \num{28.7}, \num{28.4}, \num{28.1}, \num{27.9}, \num{27.7}, and \SI{27.6}{\MHz}, respectively.
Moreover, picoRing \textit{joystick}, which can detect a center~(idle), right, up, left, and down motion of the joystick, consists of a ring base, an $8$-turn resonant coil with a joystick switch~(SKRHAAE010, Alps Alpine), and a joystick knob.
The resonant frequency of the five types of positions was tuned at \num{28.7}, \num{28.4}, \num{28.1}, \num{27.8}, and \SI{27.6}{\MHz}, respectively.
The prior rings which needs user-dependent calibration for 1-D or 2-D subtle finger input~\cite{Boldu2018FingerRader, Kienzle2014LightRing, Parizi2019AuraRing}.
However, picoRing, which directly senses the 1-D/2-D slide via the switch, can recognize the 1-D/2-D finger motions without any user-dependent calibration for each input, similar to standard physical-button-based video or game controllers.
Such intuitive operation allows for easy and quick use of picoRing \textit{slide} and picoRing \textit{joystick} for various users and situations.

\subsection{picoRing \textit{scroll}}
\label{sec:app_scroll}

Next, \autoref{fig:application}d shows picoRing \textit{scroll} used as a page controller.
picoRing \textit{scroll} consists of an inner ring that has four tiny magnets with \SI{2}{\mm} diameter and \SI{0.5}{\mm} height, an outer ring that includes an array of three reed switches~(MK24-A-2, Standex-Meder Electronics), and a bearing ring with a \SI{1.5}{\mm} width between the inner and outer rings.
Each part was 3D-printed using an SLS printer and Nylon material.
The tiny magnet placed near the reed switch can activate the reed switch when the magnet is within \SI{5}{\mm} distance.
This setup allows picoRing \textit{scroll} to recognize the rotation of the outer ring with an approximate \SI{45}{\degree} resolution, considering the two states~(\textit{i.e}, on and off states) of three reed switches. 
The resonant frequency in \textit{on} state of each reed switch was tuned at \num{29.3}, \num{28.9}, and \SI{28.6}{\MHz}.
Note that picoRing \textit{scroll} becomes relatively thick owing to the thickness of commercially-available tiny reed switch~(see \autoref{tab:ring}).
However, picoRing \textit{scroll} equipped with tiny magnets can radically reduce the risk of accidental contact with daily metallic objects, unlike the conventional ring-based scroll device equipped with bulky magnets~\cite{Ashbrook2011Nenya}.
As a result, the compact design of picoRing \textit{scroll} can enhance its usability while ensuring seamless scrolling in various applications.

\subsection{Input accuracy}
\label{sec:input_accuracy}

Lastly, we invited three participants~(one woman and two men) to evaluate how picoRing's input accuracy changes based on the different users.
Their hand size was almost similar to fit in the ring and wristband coils to their hand.
First, we measured the SNR of picoRing \textit{press}, \textit{slide}, \textit{joystick}, and \textit{scroll} around \numrange{30}{70}\si{\degree} finger bending.
\autoref{fig:application} shows the time-series SNR and the average SNR. 
The result shows that the average SNR of the four types of picoRing can be over the threshold of \SI{10}{\dB}.
Then, we attached picoRing \textit{press} to the users and measured the press recognition accuracy for the measured SNR ranging from $10$ to $13$. 
Here, we measured the recognition ratio of only picoRing \textit{press} because picoRing \textit{press}, which consists of the $7$-turn coil unlike the $8$-turn other picoRing, shows the smallest SNR.
We observed that the recognition accuracy for the pressing of $300~(=100\times \mbox{3 users})$ times was $99.7\%$ around the SNR of $11-13$, but the recognition accuracy started degrading to $93.3\%$ at the SNR of around $10$.
Such accuracy drop at the SNR of $10$ is considered negligible since the users had to bend their wrist tightly, making it hard to input to the ring. 
While a more through evaluation dedicated for each input through Fitts's law is necessary, current picoRing shows technically sufficient high SNR, ensuring reliable recognition for pressing, sliding, and scrolling actions.

\section{Discussion}
\label{sec:limit}

There are some limitations in the current picoRing.
First, the current prototype needs four types of picoRing covering various thumb-to-index inputs.
Although wearing four types of picoRing might be one of the solutions, the inductive interference between the ring coils, which has the almost similar resonant frequencies, inevitably causes wrong classification of finger inputs.
Additionally, wearing multiple rings on a single index finger could compromise comfort and wearability.
To solve this challenge, we would try to develop a tiny all-in-one switch supporting press, slide, and scroll interactions, as described in \S~\ref{sec:ring_design}, and also assign the different resonant frequencies to each interaction.
Then, picoRing requires users to attach the two types of wearable devices---ring and wristband---unlike the prior single wearables like smartwatches.
Next, our wristband device including the VNA function can potentially be miniaturized to the smartwatch size, although the current wristband requires the large VNA equipment, by referring to the S21 measurement circuit part (2 cm x 3 cm) of NanoVNA (open-source compact VNA).

Finally, the current prototype is designed for middle-sized hands.
To be available for various hand sizes, we should quantitatively explore how the coil size, the distance, and the readout frequency affect the SNR.
we also plan to examine subjective usability using System Usability Scale and NASA Task Load Index.

As for the future works, picoRing could be used to track the continuous finger movements utilizing the strength change of the inductive coupling according to the finger bending~\cite{Huang2022reconstruct, Parizi2019AuraRing}.
Also, picoRing system could support both subtle finger inputs and dynamic hand gestures by installing the conventional wristband-based hand sensing system to the wristband coil.
Moreover, the attachment of the multiple rings to both hands could provide uni-manual thumb-to-index input interface.
For example, wearing two picoRing \textit{joystick} to the left and right index fingers can enable wearable VR/AR game controller, almost similar to the standard VR/AR controllers.

\section{Conclusion}

This paper proposed picoRing, enabling a battery-free ring-based finger input device.
With the combination of the sensitive coil, the sensitive impedance measurement circuit, and the sensitive peak detection, picoRing could pair a compact wristband coil with a tiny battery-free ring coil up to \SI{13}{\cm} away from the wristband despite the finger bending and metal near the wrist.
picoRing also demonstrated that the ring coil could detect the basic subtle finger inputs including press, 1-D/2-D slide, or scroll, by using different passive switches.
The development of the single ring coil supporting various finger inputs could be one of our main future work.
We strongly believe the battery-free design of picoRing can be applied to other accessories such as earbuds, gloves, stylus, etc., and could promote the ubiquitous development of battery-free wearable input function into various accessories.

\begin{acks}
This work was mainly supported by Meta Inc. Reality Labs, partially JST ACT-X, Japan under JPMJAX21K9, and JSPS International Leading Research, Japan under 22K21343.
\end{acks}


\bibliographystyle{ACM-Reference-Format}
\bibliography{reference}

\end{document}